
\hsize  = 37 pc
\vsize = 50 pc
\rightline{UIC HEP-TH/93-4}
\rightline{April 1993}
\baselineskip = 18 true pt
\centerline {\bf New Exactly Solvable Hamiltonians:}
\centerline {\bf Shape Invariance and Self-Similarity}
\vskip 1.0 true cm
\centerline {\bf D.T. Barclay, R. Dutt$^{(a)}$, A. Gangopadhyaya$^{(b)}$,
Avinash Khare$^{(c)}$,}
\centerline {\bf A. Pagnamenta and U. Sukhatme}
\vskip 0.4 true cm
\centerline {Department of Physics, University of Illinois at Chicago,}
\centerline {Chicago, Illinois 60680, U.S.A.}
\vskip 0.8 true cm
\noindent {\bf Abstract:} We discuss in some detail
the self-similar potentials of Shabat and
Spiridonov which are reflectionless and have an infinite number of bound
states.
We demonstrate
that these self-similar potentials are in fact shape
invariant potentials within the formalism of supersymmetric quantum
mechanics. In particular, using a scaling ansatz for the change of
parameters, we obtain a large class of new, reflectionless, shape
invariant potentials of which the Shabat-Spiridonov ones are a special
case.
These new potentials can be viewed as q-deformations of the single
soliton solution corresponding to the Rosen-Morse potential.
Explicit
expressions for the energy eigenvalues, eigenfunctions and transmission
coefficients for these potentials are obtained.
We show that these potentials can also be obtained numerically.
Included as an intriguing case is a shape invariant double well
potential whose supersymmetric partner potential is only a single well.
Our class of exactly solvable Hamiltonians is further enlarged
by examining two new directions: (i) changes of parameters which are
different from the previously studied cases of translation and
scaling; (ii) extending the usual concept of shape invariance in
one step to a multi-step situation.
These extensions can be viewed as q-deformations of the harmonic oscillator or
multi-soliton solutions corresponding to the Rosen-Morse potential.
\vskip 1.5 true cm
\item {(a)} Department of Physics, Visva Bharati University,
Santiniketan 731235, India.
\item {(b)} Department of Physics, Loyola University Chicago, Chicago, IL
60626, U.S.A.
\item {(c)} Institute of Physics, Sachivalaya Marg, Bhubaneswar
751005, India.
\vfill
\eject
\noindent {\bf I. Introduction}

Recently Shabat [1] and Spiridonov [2] have discussed potentials
which are reflectionless and have an infinite number of bound states.
In addition, these potentials have the remarkable property of being
self-similar and can be looked upon as q-deformations of the
single soliton solution corresponding to the Rosen-Morse potential.
Normally, in quantum Lie algebras one takes the underlying space to be
noncommutative and the deformation parameter $q$ measures deviation
from normal analysis.
In contrast, an interesting feature of Shabat [1] and Spiridonov's
work [2] is that they have considered the same problem in a commutative
space with the q-deformation arising from the specific nature of
the potential.
Spiridonov [3] has also considered the deformation of
parasupersymmetric quantum mechanics [4] and
obtained potentials which can be regarded as a q-deformation of
the two soliton solution corresponding to the Rosen-Morse potential.

Another interesting advance of recent years has been in
supersymmetric quantum mechanics [5], where new
insight into exactly solvable potentials has been obtained
through the concept of shape invariance.
It has been shown [6] that
for supersymmetric partner potentials $V_{\pm}(x,a_0)$ satisfying
the properties of shape invariance and unbroken supersymmetry, one can
write down the energy eigenvalues algebraically.
Subsequently it was shown that both the eigenfunctions [7] and the
scattering matrix [8] can also be obtained algebraically for these
potentials.
The shape invariance condition is given by
$$V_+(x,a_0)=V_-(x,a_1)+R(a_0),   \eqno{(1.1)}$$
where $a_0$ is a set of parameters, $a_1=f(a_0)$ is an arbitrary
function describing the change of parameters and the remainder
$R(a_0)$ is independent of $x$. Certain solutions to the shape
invariance condition are known [9] including essentially all the
standard problems discussed in quantum mechanics textbooks.
In all these  cases $a_1$ and $a_0$ have been related
by a translation $(a_1=a_0+\alpha)$.
Careful analyses with this ansatz have failed to yield any additional
shape invariant potentials [10].
Indeed it has been suggested [11] that there are no other shape
invariant potentials.
Although a rigorous proof has never been given, no counter examples
have so far been found either.

In this paper we show that the Shabat-Spiridonov
(SS) self-similar potentials can be understood within the framework of
shape invariance.
In particular, we show that by using a scaling ansatz $(a_1=qa_0)$
for the change of parameter $a_0$, a large class of new shape invariant
potentials which are reflectionless and possess an infinite number of
bound states can be obtained [12].\footnote*{This is slightly
misleading in that a reparameterisation of the form $a_1=qa_0$ can
be recast as ${a'}_1={a'}_0+\alpha$ merely by taking logarithms.
However, since the choice of parameter is usually an integral part
of constructing a shape invariant potential, it is in practice part
of the ansatz. For instance, in section III, we will construct
potentials by expanding in $a_0$, a procedure whose legitimacy
and outcome are clearly dependent on our choice of parameter (and
hence reparameterisation). Note that, although the construction
is non-invariant, the resulting potentials will still be invariant
under redefinitions of $a_0$.}
Our potentials contain the self-similar potentials
of Shabat [1] and Spiridonov [2], but are considerably more general.

The plan of the paper is the following. In Section II, we discuss in some
detail the self-similar potentials of Shabat and Spiridonov (SS).
An unfortunate feature of these potentials is that they are not
known in analytical form for all $x$ and
we therefore give graphs of these potentials for some values of the
deformation parameter, which we denote by $p$ ($0<p<1$).

In Section III, we briefly discuss the shape invariance condition within
the formalism of supersymmetric quantum mechanics. Using a scaling
ansatz $(a_1=qa_0)$, we
obtain a large class of new reflectionless shape invariant potentials.
Explicit expressions for the eigenvalues, eigenfunctions and transmission
coefficients for these potentials are derived.
These potentials can be viewed as
q-deformations of a one dimensional harmonic oscillator or of the
single soliton solution corresponding to the Rosen-Morse potential.
The self-similar potentials of SS [1,2] are rederived as a special case.

In Section IV, we discuss a new technique that essentially solves
the inverse scattering problem for this wider class numerically
and so enables us to calculate these potentials.
Examples of the results obtained are discussed.

In Section V, we give the Taylor series expansion of the potentials
for large $x$.
By using these reflectionless potentials as solutions of the KdV equation,
we then estimate the area under them and indicate how one can
also estimate higher moments for these cases rigorously even though
the potentials are not known in analytical form. Using the ground state
wave function for these potentials, we also give graphs of a
continuous parameter family of potentials which are strictly
isospectral to one of the self-similar potentials.

Section VI introduces multi-step scaling ans\"atze for the change of
parameters and hence obtains new shape invariant potentials
which can be looked upon as q-deformations of the multi-soliton
solutions corresponding to the Rosen-Morse potential.

In Section VII, we discuss various other ans\"atze for connection between
parameters $a_1$ and $a_0$ and obtain yet more new shape invariant
potentials. Explicit expressions for the eigenvalues and the eigenfunctions
of these potentials are also given.

Finally, in Section VIII, we summarize the results of this paper and
indicate some open problems.
\vskip 0.5 true cm
\noindent {\bf II. Self-Similar Potentials}

Shabat [1] and Spiridonov [2] discussed an infinite
chain of reflectionless Hamiltonians given by $(\hbar = 2m = 1)$
$$H_n=P^2+V_n(x),\qquad n = 0,1,2,\ldots\eqno{(2.1)}$$
with
$$V_0(x)= W^2_0-W'_0+C_0\eqno{(2.2)}$$
and
$$V_{n+1}(x) = V_n(x)+2 W'_n(x).   \eqno{(2.3)}$$
The various superpotentials $W_n(x)$ satisfy the following set of
differential equations
$$W^2_n+W'_n=W^2_{n+1}-W'_{n+1}+C_{n+1},\qquad  n = 0,1,2,\ldots\eqno{(2.4)}$$
where the $C_n$ are arbitrary positive constants.
It is amusing to note that relations (2.4)
arise naturally in the framework of parasupersymmetric quantum
mechanics [4]. Let us assume at this stage that all $W_n(x)$ are such
that the functions
$$\psi_0^{(n)}(x) \propto \exp \Bigl[-\int^xW_n(y) dy \Bigr],\eqno{(2.5)}$$
are square integrable, and hence
correspond to the ground state wave function of the Hamiltonian
$H_n$.
In this case one has the standard situation of unbroken supersymmetry
and the potential $V_{i+1}(x)$ has one bound state less
than $V_i(x)$. Using eqs.(2.1) to (2.4), it
follows that the eigenvalues of $H_0$ are given by
$$E_m^{(0)}=\sum_{i=0}^m C_i,\qquad m = 0,1,2,\ldots \eqno{(2.6)}$$
while the corresponding eigenfunctions are given by
$$\psi_m^{(0)}(x)\alpha
(P+iW_0)(P+iW_1).....(P+iW_{m-1})\psi^{(m)}_0(x).\eqno{(2.7)}$$
It should be noted here that the Hamiltonian $H_j$ has the same
spectrum as $H_0$ except that the lowest $j$ levels of $H_0$ are
missing.

In general, it is not possible to determine these potentials unless one
imposes some extra constraints. SS specify superpotentials by
demanding that all superpotentials  $W_n(x)$ satisfy the self-similar ansatz
$$W_i(x) = p^i W(p^i x)\eqno{(2.8)}$$
with $0<p<1$.
Thus, there is just one unknown function $W(x)$.
On using $i = 0$ and 1 in eq.(2.8) and eq.(2.4), one
obtains the following finite-difference differential equation defining
$W(x)$:
$$W^2(x)+W'(x) = p^2W^2(px)-p^2 W'(px)+C_1\eqno{(2.9)}$$
where the prime denotes differentiation with respect to the argument.
Eqs.(2.8) and (2.9) are the key statements underlying the concept
of self-similarity.
On using
$i=2,3,\ldots$ and eqs.(2.4), (2.8) and (2.9) one then concludes that
$$C_n = (p^2)^{n-1}C_1\eqno{(2.10)}$$
and hence the $m$th eigenvalue of $H_0$ is given by
$$E^{(0)}_m=C_0 + {C_1(1-p^{2m)}\over (1-p^2)},
     \qquad \quad m=0,1,2,\ldots\eqno{(2.11)}$$
One can choose the arbitrary constant $C_0$ to be zero, which corresponds
to taking $E^{(0)}_0=0$.
An alternate convenient choice is to pick $C_0$ such that
$$Lim_{m\rightarrow\infty} E^{(0)}_m = 0,\eqno{(2.12)}$$
which gives
$$E^{(0)}_m=- {C_1 p^{2m}\over (1-p^2)},\quad m = 0,1,2,\ldots\eqno{(2.13)}$$

One can try to find $W(x)$ by solving the finite-difference
differential eq.(2.9) in a Taylor series form near $x= 0$;
if
$$W(x) =\sum^{\infty}_{j=0} \quad b_j x^j\eqno{(2.14)}$$
then
$$b_1(1+p^2)+b^2_0(1-p^2)=C_1\eqno{(2.15)}$$
and
$$b_{j+1} = -{(1-p^{j+2})\over (j+1)(1+p^{j+2})} \sum^j_{m=0}
      b_m b_{j-m}, \qquad j=1,2, \ldots \eqno{(2.16)}$$
Normalizability of wave functions is ensured if $W(x)$ is a continuous
function, positive at $x\rightarrow\infty$ and negative at
$x\rightarrow-\infty$. This is the case if one chooses $b_0=0$.
In this case, it follows from (2.16) that all even coefficients
$b_{j}$ ($j=0,2,4,\ldots$) are zero and hence $W(x) = -W(-x)$.
In particular one finds
$$W(x) = {C_1\over (1+p^2)} x - {1\over 3}\Bigl({C_1\over
1+p^2}\Bigr)^2{(1-p^4)\over (1+p^4)} x^3 + 0(x^5).\eqno{(2.17)}$$

Some special cases are worth noting. At $p=1$ the solution of eq.(2.9)
is the standard one-dimensional harmonic oscillator with
$W(x)=C_1 x/2$, while
in the limit $p\rightarrow 0$, the solution of (2.9) is the
one soliton superpotential corresponding to the Rosen-Morse
potential given by
$$W(x)= \sqrt{C_1} \tanh (\sqrt{C_1}x).\eqno{(2.18)}$$
Hence the general solution to eq.(2.9) with $0<p<1$
can be looked upon as a deformation of the hyperbolic tangent function
with $p$ acting as the deformation parameter.
Notice that the number of bound states as given by (2.13) increases
discontinuously from just one at $p=0$ to infinity for $p>0$.

It is important to note that the superpotentials $W(x,C_1)$ which
solve the self-similarity condition (2.9) have the simple scaling
property
$$	W(x,C_1)= \sqrt{C_1} F( \sqrt{C_1} x).   \eqno(2.19)$$
Thus, for a particular $p$ one only needs to find $W(x,C_1)$ for
any one non-trivial value of $C_1$.
Knowing $W(x)$, one immediately knows $V_0(x)$ and the other potentials
$V_n(x)$ can be recursively obtained from $V_0(x)$ using
eqs.(2.1) to (2.4), (2.8) and (2.10), so that the whole chain of
potentials are known in principle once $V_0(x)$ is known.

Unfortunately, merely knowing the Taylor series about $x=0$ is
not sufficient if one wants to carry out this programme, since
simple arguments show this series (2.17) to have a radius of
convergence $R$, where
$$  {\pi \over 2} \le \sqrt{C_1} R \le {\pi \over 2}
	\sqrt{{1+p^2 \over 1-p^2}}, \qquad 0<p\leq 1.\eqno{(2.20)}$$
The lower bound derives from noticing that the series coefficients
are smaller for $0<p<1$ than for $p=0$, the latter being essentially
the expansion of $\tanh( \sqrt{C_1} x)$ and known to have radius
of convergence $\pi/2$ (see also [13]).
The upper bound involves realising that there still has to be a pole
on the imaginary axis when $p>0$: from (2.17) one sees that along
that axis $W(ix)=i \omega(x)$ inside any radius of convergence, where
$\omega(x)$ is a real function satisfying
$$ -\omega^2(x)+ \omega'(x)= -p^2\omega^2(px)-p^2\omega'(px)+C_1, \eqno(2.21)$$
but also that $\omega(x) > \omega(px)$ and $\omega'(x) > \omega'(px)$, so
that
$$  \omega'(x)(1+p^2) > \omega^2(x)(1-p^2)+ C_1   \eqno(2.22)$$
implying that $\omega(x)$ grows faster than $\tan( \sqrt{C_1}
\sqrt{1-p^2}x/\sqrt{1+p^2})$ and so has the requisite singularity.

In the absence of a solution to (2.9) in terms of elementary
functions for all $p$, one must resort to some sort of numerical
determination other than by trying to sum the Taylor series.
The most direct approach relies on noticing that since $px<x$, if
one already knows $W(x)$ in an interval, then one knows the right hand side
of (2.9) in a larger interval.
This allows us to treat the right hand side as an already known function of $x$
and thus integrate up the equation using a (fourth order) Runge-Kutta
method.
To start this process off, the Taylor series was summed to 75 terms
on an interval well inside the radius of convergence.
The superpotentials thus obtained have been checked both by comparing
them to the summed series throughout the region where the latter is still
valid and by direct numerical integration of the Schr\"odinger
equation with the spectra checked against (2.13).
In both cases the degree of agreement is extremely high.
Examples of the kind of superpotentials and potentials obtained
are shown in Figures 1 and 2.
Note that, having independently calculated these functions, we see
no evidence of the oscillations in them reported by in [13].

Aside from its practical utility, the insight underlying the numerical
determination also strengthens confidence that (2.9) actually has
a solution.
Because any solution in a finite interval can be analytically continued to
arbitrarily large $x$, the question of existence clearly reduces
to establishing it in a neighbourhood around $x=0$.
But this is precisely the place where a convergent Taylor series
is known to exist and this is sufficient.

Additional analytic properties of these potentials will be
described in Section V, but first we wish to introduce our
more general class of potentials to which the methods of that
section will also apply.
\vskip 0.5 true cm
\noindent {\bf III. Shape Invariance With Scaling Ansatz}

A fresh impetus to the study of exactly solvable problems in
nonrelativistic quantum mechanics was provided by Gendenshte\u \i n [6]
with the introduction of shape invariant partner potentials within
the framework of supersymmetric quantum mechanics.
To set our notations, we give a quick review of both supersymmetric
quantum mechanics  and shape invariance [9].
The partner Hamiltonians $H_{\pm}$ are given by
$$H_- = A^+ A,\qquad H_+ = AA^+,\eqno{(3.1)}$$
where $(\hbar = 2m = 1)$
$$A = {d\over dx} +W(x), \qquad A^+ ={d\over dx}-W(x), \eqno{(3.2)}$$
so that the two partner potentials $V_{\pm}(x)$ can be expressed in
terms of the superpotential $W(x)$ thus
$$V_{\pm}(x) = W^2(x) \pm  W'(x).\eqno{(3.3)}$$
{}From here it follows that all the energy eigenvalues of $H_{\pm}$ are
positive semidefinite.
Further, it turns out that in case SUSY is unbroken the ground state
energy of one of the two Hamiltonians is zero and all other energy
eigenvalues of $H_{\pm}$ are paired
$$E^{(-)}_0=0,\qquad E^{(-)}_{n+1} = E^{(+)}_n.\eqno{(3.4)}$$
Here, for convention's sake, we always consider the situation of unbroken
SUSY and so the ground state energy of $H_-$ is zero.
The corresponding eigenfunction $\psi^{(-)}_0(x)$ (which satisfies
$A\psi^{(-)}_0(x)=0)$ turns out to be
$$\psi^{(-)}_0(x) = N e^{-\int^x W(y)dy}.\eqno{(3.5)}$$
One can also show that because of SUSY the eigenfunctions and
scattering amplitudes of the two partner Hamiltonians are also related
$$\psi^{(+)}_n =A\psi^{(-)}_{n+1}/\sqrt{E^{(+)}_n},
 \qquad \psi^{(-)}_{n+1} =
A^+\psi^{(+)}_n/\sqrt{E^{(+)}_n},\eqno{(3.6)}$$
$$R_-(k)={W_-+ik\over W_--ik}R_+(k),\eqno{(3.7)}$$
$$T_-(k)={W_+-ik'\over W_--ik}T_+(k),\eqno{(3.8)}$$
where $k = (E-W_-^2)^{1/2}$ and $k' =(E-W^2_+)^{1/2}$ with
$W_{\pm} =W(x=\pm\infty)$.

If the pair of SUSY partner potentials $V_{\pm}(x)$
defined by eq.(3.3) differ only via the parameters that appear in them,
then they are said to be shape invariant [6]; that is, if the partner
potentials $V_{\pm}(x,a_0)$ satisfy the condition (1.1).
In terms of the superpotential $W$, this shape invariance condition reads
$$W^2(x,a_0)+W'(x,a_0)=W^2(x,a_1)-W'(x,a_1)+R(a_0).\eqno{(3.9)}$$
The common $x$-dependence in $V_-$ and $V_+$ allows a full
determination of energy eigenvalues [6], eigenfunctions [7] and
scattering matrices [8] algebraically. One finds
$$E^{(-)}_n(a_0)=\sum^{n-1}_{k=0} R(a_k),\qquad
E^{(-)}_0(a_0)=0,\eqno{(3.10)}$$
$$\psi^{(-)}_n(x,a_0)=A^+(x,a_0)A^+(x,a_1) \ldots
A^+(x,a_{n-1})\psi^{(-)}_0(x,a_n).\eqno{(3.11)}$$

It is still a challenging open problem to identify and classify all the
solutions to the shape invariance condition (3.9). Certain solutions to
it are known [9] and they include essentially all exactly solvable
problems discussed in standard texts on quantum mechanics. For all
these, $a_1$ and $a_0$ are related by a translation. Careful analysis
with this ansatz has failed to uncover any additional shape invariant
potentials [10] and in fact it has been suggested that there are no
others [11]. We shall now show that this is
not the case since a large number of new shape invariant
potentials can result from a new scaling ansatz
$$ a_1 = q a_0\eqno{(3.12)}$$
where $0<q<1$, a choice motivated by recent interest
in q-deformed Lie algebras.
Our approach includes the self-similar potentials
discussed in the previous section as a special case.

Consider an expansion of the superpotential of the form
$$W(x,a_0) =\sum^{\infty}_{j=0} g_j(x)a_0^j.\eqno{(3.13)}$$
Using eqs.(3.12) and (3.13) in the shape invariance condition (3.9),
writing $R(a_0)$ in the form
$$R(a_0)=\sum^{\infty}_{j=0} R_ja_0^j\eqno{(3.14)}$$
and equating powers of $a_0$ yields
$$2g'_0(x) = R_0,\qquad g'_1(x)+2d_1g_0(x)g_1(x)=r_1 d_1, \eqno{(3.15)}$$
$$g'_n(x)+2d_n g_0(x)
g_n(x)=r_nd_n-d_n\sum^{n-1}_{j=1}g_j(x)g_{n-j}(x)\eqno{(3.16)}$$
where
$$r_n\equiv R_n/(1-q^n), \quad d_n=(1-q^n)/(1+q^n),
         \quad n=1,2,3,\ldots   \eqno{(3.17)}$$
This set of linear differential equations is easily solvable in
succession to give a general solution of eq.(3.9). Let us first
consider the special case
$$g_0(x) = 0 \eqno{(3.18)}$$
which implies $R_0=0$.
The general solution of (3.16) is then
$$g_1(x)=r_1 d_1 x, \qquad
g_n(x)=d_n\int dx \Bigl[r_n-\sum^{n-1}_{j=1} g_j(x)g_{n-j}(x)\Bigr],
\qquad n=2,3, \ldots\eqno{(3.19)}$$
where without any loss of generality we have assumed all the
constants of integration to be zero.
The shape invariance condition thus essentially fixes the $g_n(x)$
(and hence $W(x,a_0)$ via (3.13)) once $R(a_0)$ is specified, i.e.
once the set of $r_n$ are chosen.
Implicit constraints on this choice are that the resulting
ground state wavefunction (3.5) be normalisable and, so that the
spectrum (3.10) is sensibly ordered, that $R(q^n a_0)>0$.

The simplest case is $r_1>0$ (positivity required to ensure
normalisable wavefunctions) and $r_n=0$, $n \geq 2$.
Here (3.19) takes on a particularly simple form
$g_n(x)= \beta_n x^{2n-1}$ with
$$ \beta_1 = d_1r_1, \qquad
\beta_n =  - {d_n \over {(2n-1)} }
 \sum_{j=1}^{n-1} \beta_j \beta_{n-j} n=2,3, \ldots
\eqno(3.20)$$
and so
$$ W(x,a_0) = \sum_{i=1}^{\infty} \beta_i a_0^i x^{2i-1}
    = \sqrt{a_0} F(\sqrt{a_0} x). \eqno(3.21)$$
For $a_1=q a_0$ this now gives
$$ W(x,a_1)= \sqrt{q} W(\sqrt{q}x, a_0),  \eqno(3.22)$$
hence in this special case the shape invariance condition (3.9)
becomes
$$ W^2(x,a_0) + W'(x,a_0) = q W^2(\sqrt{q}x, a_0)
   -q {dW \over d \sqrt{q}x} (\sqrt{q}x, a_0) + a_0r_1(1-q).  \eqno(3.23)$$
Comparing this to (2.9), one thus sees that the case $r_n=0$, $n \geq2$
corresponds to the self-similar $W$ of Shabat and Spiridonov provided
one writes $\gamma^2 \equiv d_1 r_1 a_0$ and $q \equiv p^2$.
In fact, if instead of choosing $r_n=0,n\geq 2$, any one $r_n$
(say $r_j$) is taken to be nonzero and $q^j$ is replaced by $p^2$
then one again obtains the self-similar potentials.
In these instances the results obtained from shape invariance and
self-similarity are entirely equivalent and the Shabat-Spiridonov
self-similarity condition turns out to be a special case of the
shape invariance one (3.9).

However, it is necessary to emphasize here that shape invariance is a much more
general concept than self-similarity.
For example, if we choose more than one $r_n$ to be nonzero, then
shape invariant potentials are obtained which are not
self-similar.
As an illustration, consider $r_n=0$, $n\geq 3$.
Using eq.(3.16) one can readily calculate all the $g_n(x)$, of which the
first three are
$$g_1(x)=d_1r_1x, \qquad g_2(x)= d_2r_2x-{1\over 3}d^2_1r^2_1d_2x^3,$$
$$g_3(x)=-{2\over 3}d_1r_1d_2r_2d_3x^3+{2\over 15}d^3_1r^3_1d_2d_3x^5.
                                    \eqno{(3.24)}$$
Notice that in this case $W(x)$ contains only odd powers of $x$.
This makes the potentials $V_{\pm}(x)$ symmetric in $x$ and also guarantees
unbroken SUSY.
It is convenient to define the combinations $\Gamma_1 \equiv d_1 r_1 a_0
= \gamma^2$ and $\Gamma_2 \equiv d_2 r_2 a_0^2$.
Then, the energy eigenvalues follow immediately from eq.(3.10) and (3.14)
($0<q<1$):
$$E^{(-)}_n(a_0)=\Gamma_1{(1+q)(1-q^n)\over (1-q)}
        +\Gamma_2{(1+q^2)(1-q^{2n})\over (1-q^2)},
		\qquad n=0,1,2,\ldots \eqno{(3.25)}$$
while the (unnormalized) ground state wave function is
$$\psi^{(-)}_o(x,a_0)=\exp\bigg [ -{x^2\over
2}(\Gamma_1+\Gamma_2)+{x^4\over 4}(d_2\Gamma^2_1
+2d_3\Gamma_1\Gamma_2+d_4\Gamma^2_2)+0(x^6)\bigg].  \eqno{(3.26)}$$
The excited wave functions can be recursively calculated using eq.(3.11),
though usually it is more convenient to use the relation
$$\psi^{(-)}_n(x,a_0)=A^+(x,a_0)\psi^{(-)}_{n-1}(x,a_1).\eqno{(3.27)}$$

We can also calculate the transmission coefficient of this symmetric
potential ($k=k'$) by using relation (3.8) and the fact that for this
shape invariant potential [8]
$$T_+(k,a_0)=T_-(k,a_1=qa_0).\eqno{(3.28)}$$
Repeated application of eqs.(3.8) and (3.28) gives
$$T_-(k,a_0)={[ik-W(\infty,a_0)][ik-W(\infty,a_1)]....[ik-W(\infty,a_{n-1})]\over
[ik+W(\infty,a_0)][ik+W(\infty,a_1)]....[ik+W(\infty,a_{n-1})]}T_-(k,a_n)\eqno{(3.29)}$$
where
$$W(\infty,a_j)=\sqrt{E^{(-)}_{\infty}-E^{(-)}_j}.\eqno{(3.30)}$$
Now, as $n\rightarrow\infty$, $a_n=q^na_0 \rightarrow 0$ ($0<q<1$)
and, since we have
taken $g_0(x)=0$, one gets $W(x;a_n)\rightarrow 0$. This
corresponds to a free particle, for which
the transmission coefficient is unity; as a result
the reflection coefficient $R_-(x;a_0)$, vanishes and the
transmission coefficient is given by
$$T_-(k,a_0)=\prod^{\infty}_{j=0}{\bigg [ik-W(\infty,a_j)\bigg ]\over
\bigg [ik+W(\infty,a_j)\bigg ]}.  \eqno{(3.31)}$$
Note that this is a general expression for $T$ derived using only
$a_n=q^na_0$ and the fact that $V_-(x)$ is a symmetric potential ($k=k'$). In
particular, no assumption has been made regarding the values of $r_n$.
In the case when only one of the $r_n$ (say $r_1$) is non-zero, eq.
(3.31) gives the transmission coefficient of the
self-similar potentials (3.13) to (3.16).

The above discussion keeping only $r_1,r_2\not = 0$ can readily be
generalized to an arbitrary number of nonzero $r_j$. The energy
eigenvalues for this case are given by $(\Gamma_j \equiv d_jr_ja^j_0)$
$$E^{(-)}_n(a_0)=\sum_j \Gamma_j{(1+q^j)(1-q^{nj})\over (1-q^j)},
  \qquad n=0,1,2,\ldots      \eqno(3.32)$$
All these potentials are also symmetric and reflectionless with $T_-$
given by eqs.(3.31).

The limit $q\rightarrow 0$ is particularly simple and again yields the
one-soliton Rosen-Morse potential with $W=\alpha \tanh \alpha x$.
Thus results corresponding to different choices of $R_n$
can be regarded as multiparameter deformations of this potential.

Finally, let us consider the solution to the shape invariance condition
(3.9) in the case where $R_0$ is nonzero, so that $g_0(x)={1 \over 2}
R_0x$ from (3.15) rather than being zero.
One can again solve the set of linear differential eqs.(3.16) in succession
using this $g_0(x)$ and hence obtain $g_1(x),g_2(x),\ldots$
Further, the spectrum can be immediately found using eqs.(3.10) and (3.14);
for example, in the case of an arbitrary number of nonzero
$R_j$ (in addition to $R_0$), the spectrum is given by
$$E_n=nR_0+\sum_j\Gamma_j{(1+q^j)(1-q^{nj})\over (1-q^j)}\eqno{(3.33)}$$
which is the spectrum of a q-deformed harmonic oscillator [14].
It should be noted here that, unlike the usual q-oscillator where the space
is noncommutative, but the potential is normal $(w^2x^2)$, in our
approach the space is commutative, but the potential is
deformed, giving rise to such a multi-parameter deformed oscillator spectrum.
\vskip 0.5 true cm
\noindent {\bf IV. Numerical Results}

Explicit determination of the SS potentials (described in
Section II) crucially depended on the scaling property
$$      W(x,a_0)= \sqrt{a_0} F(\sqrt{a_0}x),   \eqno(4.1)$$
displayed by the solutions, which allowed $W'(x,a_0)$ to be
related to $W(\sqrt{q}x,a_0)$ instead of merely $W(x,a_1)
=W(x,qa_0)$.
However, such scaling is not a property of the solutions
in Section III when more than one $r_n$ is non-zero, as can be
seen from the series expansion (3.24).
When only $r_1$ and $r_2$ are non-zero, there is a generalisation
of the form $W(x,a_0)=\sqrt{r_1 a_0} F(\sqrt{r_1 a_0} x, r_2 a_0/r_1)$
which relates the behaviour at $x$ to that of {\it another problem}
(that corresponding to calculating $W(x,a_0)$ with a different $r_2$)
at $\sqrt{q}x$.
By forming a ladder of these potentials (in which $r_2$ is tending to
zero and hence the problem towards the special SS case), it should be
possible to determine $W(x,a_0)$ using this fact.
However, we chose to devote this section to a method that emphasizes
$W(x,a)$ as a function of both $x$ and $a$ and which more readily
generalises to arbitrary $r_n$.

Intuitively one would still expect the series (3.13) to be convergent
for either $x$ or $a$ sufficiently small.
Thus in the $(x,a)$ plane we can assume that $W(x,a)$ is calculable
to arbitrary accuracy close to either axis and the problem
reduces to continuing this knowledge out into the plane.
Defining sum and difference functions
$$\eqalignno{ S(x,a) &\equiv W(x,a)+W(x, qa)    &(4.2a)\cr
	      D(x,a) &\equiv W(x,a)-W(x,qa),    &(4.2b)\cr}$$
for $a_1=qa_0$, the shape invariance condition (3.10) becomes
$$ {dS \over dx} = - S(x,a)D(x,a) + R(a)  \eqno(4.3)$$
where $R(a)$ is a known function.
Now if one knows $W(x,a)$ for $x<X$, $a<A$, an Euler step
(since $D$ is only known in $x<X$, there's not enough
information available for that to be a Runge-Kutta one) will
give $S(X+h,a)$, again for $a<A$.
One must now invert (4.2a) to convert this knowledge of
$S(X+h,a)$ into information about $W(X+h,a)$.
Iterated use of (4.2a) relates $W(X+h,a)$ to $W(X+h,q^na)$
and $n$ known values of $S$.
For some sufficiently large $n$, $W(X+h,q^na)$ can be calculated
using the Taylor series (3.14) and so one can indeed determine
$W(X+h,a)$ for all $a<A$.
Breaking the $(x,a)$ plane up into a grid and using some suitable
interpolation method for the points in between, one can iterate
this Euler step up through $x$ and so obtain $W(x,a)$ numerically
for values of $x$ and $a$ limited by computing constraints only.
The only inputs are $R(a)$ and the series approximations for small
$x$ and $a$.

A program to implement this scheme has been developed and its results
for the special SS case were shown to agree very well with
the earlier (more accurate, but overly specialised) program.
As an example of the new potentials this permits us to
consider the potential corresponding to $r_1=1$, $r_2 =-1$ with all
other $r_n=0$.
Provided $a<1/(1+q)$ the spectrum given by (3.25) is
well-ordered.
In Figure 3 we display the potential calculated for
$q=0.3$ and
$a=0.75$, along with its partner potential and the exact spectrum
found from (3.25). (These eigenvalues have also been checked numerically).
Note that for this choice of parameters, $V_-(x)$ is a double well
potential, whereas its shape invariant partner $V_+(x)$ is a single
well. This situation is
unlike the (non-shape-invariant) examples discussed in Ref. [15], where the
SUSY partner of the initial double-well potential has a sharp
$\delta$-like spike at its center.
Apart from being the first shape invariant double-well, this example
stretches na\"\i ve intuition concerning shape invariance.
Furthermore, this example barely indicates the variety of behaviour available
by altering $q$, $a$ and the $r_n$ in this new class of potentials.

Finally, we note that the basic idea of this section can be divorced
from the details of the Taylor series.
The restriction to symmetric potentials gives $W(0,a)=0$, to be used
as a boundary condition for the intial Euler step.
To invert (4.2a) one can use the infinite series
$$ W(x,a)=S(x,a) - S(x,qa) + S(x,q^2a)-S(x,q^3a)+ \ldots \eqno(4.4)$$
which is convergent provided $W(x,a) \rightarrow 0$ as $a \rightarrow 0$.
More useful numerically is the fact that, because it is alternating, this
series can be truncated with a rigorous bound on the error and without
needing  to calculate some $W(X+h, q^n a)$ by use of a Taylor series.
Neither is the reliance on $a_1=q a_0$ terribly restrictive: one can
always redefine the parameters to obtain this.
The only crucial constraint is that the method applies exclusively to
symmetric potentials holding infinitely many bound states and corresponding
to a chain of Hamiltonians ($H_-$, $H_+$ etc.) which tends asymptotically
towards a free-particle one (i.e. $W(x,a_n) \rightarrow 0$ as $n
\rightarrow \infty$).
Otherwise the sole input is $R(a)$ (expressed in terms of the appropriately
defined $a$), which in general should be deducible from any desired
(possible) spectrum.
\vskip 0.5 true cm
\noindent {\bf V. Analytic Results}

Although undeniably useful, simply being able to calculate $W$ is an
unsatisfactory state of affairs unless there is also a body of
complementary analytic results.
This section therefore brings together several approaches by
which such results can be gathered.

In our method of constructing the potentials, the Taylor series
about $x=0$ played an essential role (as it did in [1] and [2]).
However, to get a better insight into the potentials,
it may be worthwhile to also know the Taylor series of
$W(x)$ around $x\rightarrow\infty$.
For simplicity, we now restrict our attention to
the SS family
and return to using the original parameter $p=\sqrt{q}$.
Substituting
$t=1/x$ in eq.(2.9) yields
$$W^2(t)-t^2W'(t)=p^2W^2(t/p)+t^2W'(t/p)+C_1.\eqno{(5.1)}$$
On substituting\footnote*{We assume, in the absence of evidence
to the contrary, that the simple Taylor series is the appropriate
expansion.}
$$W(t) = \sum^{\infty}_{j=0}\quad a_jt^j\eqno{(5.2)}$$
in this equation one obtains
$$a^2_0=C_1/(1-p^2),\qquad a_1=0\eqno{(5.3)}$$
and
$$ a_j= -(j-1)\biggl({1+p^{j-2}\over 1-p^{j-2}}\biggr){a_{j-1} \over 2 a_0}
       -{1 \over 2 a_0} \sum_{m=2}^{j-2} a_m a_{j-m},
		\qquad j=3,4,\ldots\eqno{(5.4)}$$
Thus we find that as $x\rightarrow + \infty$
$$W(x)=\sqrt{{C_1\over (1-p^2)}} + {a_2\over x^2} - {a_2\over a_0}
\biggl({1+p\over 1-p}\biggr) {1 \over x^3}+ \ldots \eqno{(5.5)}$$
where $a_2$ is an arbitrary constant. This arbitrariness is due to the
fact that $W(0)=0$ has not been imposed while deriving (5.5). In fact
it is not easy to do so since the series (5.5) is valid for large $x$.

It has already been established that we are dealing with reflectionless,
symmetric potentials for which the infinite spectra of eigenvalues are
known exactly in closed form.
This type of problem has already been well studied, but the standard
inverse scattering method [16] has proved too cumbersome to be of
much practical use in deriving these potentials in this non-trivial
context.
However, certain well-known, related results can be used to quite
strongly constrain the potentials: the point is, being reflectionless,
these can be regarded as a solution of the KdV equation at time $t=0$ [17].
Now it is known that such a solution as $t\rightarrow\pm\infty$ will
break up into an infinite number of solitons of the form $2k^2_i \hbox{sech}^2
k_ix$. On using the fact that KdV solitons obey an infinite number of
conservation laws corresponding to mass, momentum,
energy etc., one can immediately obtain constraints on the
reflectionless potentials by using the known solutions at
$t\rightarrow\pm\infty$. For example, the first three conservation
laws are
$$\int\limits_{-\infty}^{\infty}V_0(x)dx =
\sum^{\infty}_{i=0}\int\limits^{\infty}_{-\infty} V^{(i)} (x)dx\eqno{(5.6)}$$
$$\int\limits_{-\infty}^{\infty}V_0^2(x)dx =
\sum^{\infty}_{i=0}\int\limits^{\infty}_{-\infty} [V^{(i)}
(x)]^2dx\eqno{(5.7)}$$
$$\int\limits_{-\infty}^{\infty}\Bigl[V_0^3(x)+{1\over 2}\Bigl({dV_0\over
dx}\Bigr)^2\Bigr]dx = \sum^{\infty}_{i=0}\int\limits^{\infty}_{-\infty}
\Bigl[[V^{(i)} (x)]^3 +{1\over 2}({dV^{(i)}(x)\over
dx})^2\Bigr]dx\eqno{(5.8)}$$
where
$$V^i(x) = - 2 k^2_i \hbox{sech}^2 k_ix\eqno{(5.9)}$$
with $k_i=k_0 p^i$ and $k_0^2=C_1/(1-p^2)$.
Using eq.(5.9) it is straightforward to evaluate the right hand sides of
eqs.(5.6) to (5.8) and we find
$$\int\limits_{-\infty}^{\infty}V_0(x)dx =
-4\sum^{\infty}_{i=0}k_i={-4k_0\over 1-p}\eqno{(5.10)}$$
$$\int\limits_{-\infty}^{\infty}V^2_0(x)dx =
{16\over 3}\sum^{\infty}_{i=0} k^3_i={16 \over 3}{k^3_0\over
(1-p^3)}\eqno{(5.11)}$$
$$\int\limits_{-\infty}^{\infty}\Bigl[V_0^3(x)+{1\over 2}\Bigl({dV_0\over
dx}\Bigl)^2\Bigr]=-{32\over 5}\sum^{\infty}_{i=0}k^5_i=-{32\over 5}{k^5_0\over
(1-p^5)}\eqno{(5.12)}$$
thereby providing strong constraints on the potential $V_0(x)$.

All deformations of potentials considered so far have been such that
the spectra
obtained are $q$ (or $p$) dependendent.
Before ending this section, it is worth remarking that, as with any
potential, there are also distortions of the $V_n(x)$, with deformation
parameter $\lambda$, which leave the spectra unchanged. Using the techniques
of supersymmetric quantum mechanics, one can construct a large class
of strictly isospectral potentials.
For example, using any one of the
$W_i(x)$ as given by eqs.(2.5), (2.8) and (2.14) to (2.17) one can
immediately obtain a one parameter family of strictly isospectral
reflectionless potentials $V_n(x,\lambda)$ by using the formula [8]
$$V_n(x,\lambda)=V_n(x) - 2 {d^2\over dx^2} \ln (I_n(x)+\lambda)\eqno{(5.13)}$$
where $V_n(x)$ can easily be obtained using eqs.(2.1) to (2.3),
(2.8) and (2.14) to (2.17), $\lambda$ is any arbitrary parameter
($\lambda > 0$ or $\lambda < -1$) and
$$I_n(x)=\int\limits^x_{-\infty}\![\psi^{(n)}_0(y)]^2 dy.\eqno{(5.14)}$$
Here $\psi^{(n)}_0$ is as given by eq.(2.5) which can be explicitly
obtained by using eqs.(2.8) and (2.14) to (2.17). As an
illustration, we give graphs of the $V_0(x,\lambda)$ obtained from the
SS potential with $p=0.5$ for various values of $\lambda$ in Figure 4.
Large values of $\lambda$ correspond to the original SS potential
(see Fig.4c).
As $\lambda$ takes on values closer to zero, the potential gradually
breaks into two pieces, one corresponding to the $E=0$ state only
and the other containing the remaining energy levels [8].
\vskip 0.5 true cm
\noindent {\bf VI. Shape Invariance in More Than One Step}

Having obtained potentials which are multiparameter deformations
of the one soliton solution of the Rosen-Morse potential, an
obvious question to ask is if one can also obtain deformations
of the multi-soliton solutions.
The answer is yes and as an illustration we now explicitly obtain
multiparameter deformations of the two soliton case.
The desired deformation is achieved by extending the usual shape
invariance ideas to the more general concept of shape invariance
in two steps.

Consider the unbroken SUSY case of two superpotentials $W_0(x,a_0)$
and $W_1(x,a_0)$ such that $V_0^{(+)}(x,a_0)$ and $V_1^{(-)}(x,a_0)$
are the same up to an additive constant.
$$ V_0^{(+)}(x,a_0) = V_1^{(-)}(x,a_0)+R(a_0),   \eqno(6.1)$$
or equivalently,
$$ W_0^2(x,a_0)+W'_0(x,a_0)=W_1^2(x,a_0)-W'_1(x,a_0)+R(a_0). \eqno(6.2)$$
Shape invariance in two steps means that
$$ V_1^{(+)}(x,a_0) = V_0^{(-)}(x,a_1) + \tilde R(a_0),  \eqno(6.3)$$
that is
$$ W_1^2(x,a_0)+W'_1(x,a_0)= W_0^2(x,a_1)-W'_0(x,a_1)+\tilde R(a_0).
   \eqno(6.4)$$
For the above situation, the energy eigenvalues and eigenfunctions
of the potential $V_0^{(-)}(x,a_0)$ can be algebraically calculated
as shown below.

Unbroken SUSY implies zero energy ground states for the potentials
$V_0^{(-)}(x,a_0)$ and $V_1^{(-)}(x,a_0)$:
$$ E_0^{(-)0}=0, \qquad E_0^{(-)1}=0.  \eqno(6.5)$$
The degeneracy of energy levels for supersymmetric partner potentials
yields
$$ E_n^{(+)0}(a_0)=E_{n+1}^{(-)0}(a_0), \qquad
   E_n^{(+)1}(a_0)=E_{n+1}^{(-)1}(a_0). \eqno(6.6)$$
{}From eq.(6.1) it follows that
$$ E_n^{(+)0}(a_0) =E_n^{(-)1}(a_0)+ R(a_0).  \eqno(6.7)$$
For the special case $n=0$, eqs.(6.6) and (6.7) give
$$ E_1^{(-)0}=R(a_0).   \eqno(6.8)$$
Also, the shape invariance constraint (6.3) gives
$$ E_n^{(+)1}(a_0) = E_n^{(-)0}(a_1) + \tilde R(a_0). \eqno(6.9)$$
Using eqs.(6.6), (6.7), (6.9) and some algebra, one gets
$$ E_{n+1}^{(-)0}(a_0)= E_{n-1}^{(-)0}(a_1)+ R(a_0)+ \tilde R(a_0).
   \eqno(6.10)$$
These equations can be solved recursively to get
$$\eqalignno{ E_{2n}^{(-)0} &= \sum_{k=0}^{n-1}\bigl[ R(a_k) +\tilde R(a_k)
\bigr],\cr
      E_{2n+1}^{(-)0} &= \sum_{k=0}^{n-1} \bigl[ R(a_k) +\tilde R(a_k)\bigr]
         +R(a_n).  &(6.11)\cr}$$
The above discussion has been completely general and is valid for any
change of parameters, $a_1=f(a_0)$.
Following the treatment of Section III, we now take the scaling ansatz
$a_1=qa_0$ and expand the superpotentials $W_0$ and $W_1$ in powers
of $a_0$.
$$W_0(x,a_0)=\sum^{\infty}_{j=0} \quad g_j(x)a_0^j\eqno{(6.12)}$$
$$W_1(x,a_0)=\sum^{\infty}_{j=0} \quad h_j(x)a_0^j.\eqno{(6.13)}$$
Further, write $R$ and $\tilde R$ in the form
$$R(a_0) = \sum^{\infty}_{j=0}R_ja^j_0, \qquad\tilde R(a_0) =
\sum^{\infty}_{j=0}\tilde R_ja^j_0.\eqno{(6.14)}$$
Using these in eqs.(6.2) and (6.4) and equating powers of $a_0$ yields
($n=0$,1,2,$\ldots$)
$$g'_n+\sum^n_{j=0}g_j g_{n-j}=\sum^{n}_{j=0}h_jh_{n-j}-h'_n+R_n\eqno{(6.15)}$$
$$h'_n+\sum^n_{j=0}h_j h_{n-j}=q^n\sum^{n}_{j=0}g_j
g_{n-j}-q^n g'_n+\tilde R_n.\eqno{(6.16)}$$
This set of linear differential equations is easily solvable in
succession.
Let us first discuss the special case
$$g_0(x) = h_0(x)= 0,\eqno{(6.17)}$$
which implies that $R_0=\tilde R_0=0$, and further assume that
$R_n=\tilde  R_n=0$,$ n\geq 3$. In this case one can readily calculate
all $g_n(x)$ and $h_n(x)$; the first two of each are
$$g_1={(R_1-\tilde R_1)\over (1-q)}x,\quad g_2=
 {(R_2-\tilde R_2)\over (1-q^2)}x
+{x^3\over
3(1-q)^3}\bigg [ (1-q)(\tilde R^2_1-R^2_1)-2(1+q)R_1\tilde
R_1\bigg ],$$
$$h_1={(\tilde R_1-qR_1)\over (1-q)}x,$$
$$h_2= {R_2 x \over (1-q^2)}
-{x^3\over3(1+q)(1-q)^2}
\bigg [ (1+q)\tilde R^2_1+(1+q)(1-q^2)R^2_1-2q(1-q)R_1\tilde
R_1\bigg ]. \eqno(6.18)$$
It may be noted that both $W_0$ and $W_1$ contain only odd powers of
$x$ so that the potentials $V_0^{(\pm)}$ and $V_1^{(\pm)}$ are
all symmetric in $x$ and SUSY is unbroken.
The energy eigenvalues can be obtained from eqs.(6.8) and (6.11).
$$E^{(-)0}_1(a_0)=R_1a_0+R_2a^2_0,\eqno{(6.19)}$$
$$E^{(-)0}_{2n}(a_0)=\sum^2_{j=1}(R_j+\tilde R_j)a^j_0\Bigl({1-q^{jn}\over
1-q^j}\Bigr),\eqno{(6.20)}$$
and
$$E^{(-)0}_{2n+1}(a_0)=\sum^2_{j=1}R_j a^j_0\Bigl({1-q^{j(n+1)}\over
1-q^j}\Bigr)+\sum^2_{j=1} \tilde R_ja^j_0 \Bigl({1-q^{jn}\over
1-q^j}\Bigr). \eqno{(6.21)}$$
For the special case when $R_2=\tilde R_2=0$, the spectrum has been
obtained previously by Spiridonov from consideration of
self-similar potentials [3]. However, the spectrum in the general
case given by eqs.(6.20) and (6.21) cannot be obtained in such a fashion.
The energy eigenvalues
$E^{(-)1}_n$ are now immediately obtained from eq.(6.9) and the energy
eigenfunctions and transmission coefficient for these
reflectionless potentials can also be found using
eqs.(3.5), (3.27), (3.30) and (3.31). Further, the above discussion
can be readily generalized to an arbitrary number of nonzero
$R_j,\tilde R_j$.

The limit $q\rightarrow 0$ of the above equations is particularly simple
and yields the two soliton solution of the
Rosen-Morse potential, i.e.
$$W_0=2\sqrt{\tilde R} \tanh \sqrt{\tilde R} x,
	\qquad W_1=\sqrt{\tilde R} \tanh \sqrt{\tilde R} x\eqno{(6.22)}$$
provided $R=3\tilde R$. Thus our results can be regarded as
multi-parameter deformations of this potential.

Finally, it is clear that one can easily generalize this procedure
and consider shape invariance with a scaling ansatz in 3,4,$\ldots p$ steps
and thereby obtain multi-parameter deformations of the
3,4,$\ldots p$ soliton Rosen-Morse solution.
\vskip 0.5 true cm
\noindent {\bf VII. Shape Invariance with a Non-Scaling Change
of Parameters}

We have so far obtained new shape invariant potentials for $a_1$ and
$a_0$ related by the scaling ansatz $(a_1=qa_0)$. Are there shape
invariant potentials where $a_1$ and $a_0$ are neither related by
scaling  nor by translation $(a_1=a_0+\alpha)$?
We now demonstrate the existence of yet other possibilities by
obtaining potentials for $a_1=qa_0^p$ and $a_1=qa_0/(1+ pa_0)$.

First consider the case when
$$a_1=qa_0^p\eqno{(7.1)}$$
where $p$ could be any integer.
Again consider the expansions of the superpotential $W$ and
$R(a_0)$ given by eqs.(3.13) and (3.14) respectively. On using
eqs.(3.12), (3.13) and (7.1) in the shape invariance condition
(3.9) and equating powers of $a_0$ one finds two sets of equations
\item {(i)} $n = pm$, $m = 0,1,2,\ldots$
$$g'_{pm}(x)+\sum^{pm}_{j=0}g_j(x)g_{pm-j}(x)=q^m\sum^m_{j=0}g_j(x)g_{m-j}(x)-q^m
g'_m(x)+R_{pm}\eqno{(7.2)}$$
\item {(ii)} $n = pm + q$, $q = 1,2 \ldots (p-1)$
$$g'_{pm+q}+\sum^{pm+q}_{j=0}g_j(x)g_{pm+q-j}(x)=R_{pm+q}.\eqno{(7.3)}$$

These sets of equations are easily solved in
succession to produce more solutions of eq.(3.9). Further, the
energy eigenvalue spectrum can be easily obtained from eqs.(3.10),
(3.14) and (7.1). For example, in the case where only $R_1$ and $R_2$ are
nonzero the spectrum can be shown to be ($E^{(-)}_0=0$)
$$E^{(-)}_n={R_1\over q^{1/(p-1)}}\sum^n_{j=1}
	(q^{{1\over p-1}}a_0)^{p^{(j-1)}} + {R_2\over q^{2/(p-1)}}
	\sum^n_{j=1}(q^{{1\over p-1}}a_0)^{2p^{(j-1)}},\qquad n=1,2,\ldots
\eqno(7.4)$$
The energy eigenfunctions and the transmission coefficient for these
reflectionless potentials can be written down immediately using
eqs.(3.5), (3.27), (3.30) and (3.31).

As an illustration, let us discuss the case $p = 2$ explicitly.
The set of equations which follows from eqs.(7.2) and (7.3) is
$$g'_{2m}(x)+\sum^{2m}_{j=0}g_j(x)g_{2m-j}(x)=q^m\sum^m_{j=0}g_j(x)g_{m-j}(x)-q^m
g'_m(x)+R_{2m}\eqno{(7.5)}$$
$$q'_{2m+1}(x)+\sum^{2m+1}_{j=0}g_j(x)g_{2m+1-j}(x)=R_{2m+1}\eqno{(7.6)}$$
and one can thus readily calculate all the $g_n(x)$. For example, in the
case when only $R_1$ and $R_2$ are nonzero it is easily shown that
the first three $g(x)$'s are
$$g_1(x)=R_1 x,\qquad  g_2(x) = (R_2-qR_1)x -{1\over 3} R_1^2 x^3,$$
$$g_3(x)={2\over 3}R_1(qR_1-R_2)x^3 +{2\over 15} R^3_1 x^5.\eqno{(7.7)}$$
Notice that we have chosen $g_0(x)=0$ so that again $W(x)$ contains only odd
powers of $x$, $V_{\pm}(x)$ are symmetric in $x$ and SUSY is unbroken.
The spectrum which follows from (7.4) (for $p=2$) is $E^{-}_0=0$ and
$$E^{(-)}_n={R_1\over q}\sum^n_{j=1}(a_0q)^{2^{j-1}}+{R_2\over q^2}
	\sum^n_{j=1}(a_0q)^{2^j}, \qquad n=1,2,\ldots \eqno{(7.8)}$$
The $q\rightarrow 0$ limits of the equations above again correspond to the
one soliton solution of the Rosen-Morse potential, so
that our results for $a_1=qa_0^p$ can be regarded as multi-parameter
deformations of this potential. Generalization to the case when an
arbitrary number of $R_j$ are nonzero is straightforward. Similarly,
one can also consider shape invariance in multi-steps along with the
ansatz (7.1), thereby obtaining deformations of the multi-soliton
solutions.

Finally, consider solutions to the shape invariance
condition (3.9) for
$$ a_1={qa_0\over 1+pa_0}\eqno{(7.9)}$$
where $0 < q$, $ p < 1$. We also assume that $pa_0 < < 1$ so that one
can expand $(1+pa_0)^{-1}$ in powers of $a_0$. Further, assume
that in eqs.(3.12) and (3.13)
$$R(a_0) = R_1a_0+R_2a^2_0\eqno{(7.10)}$$
and $g_0(x)=0$ so that $W$ is again an odd function of $x$.
On using eqs.(3.12), (3.13), (7.9) and (7.10) in the  shape
invariance condition (3.9), expanding negative powers of $(1+pa_0)$
in powers of $a_0$ and finally equating powers of $a_0$, one again
obtains a set of linear differential equations.
For example, to order $a^2_0$, the shape invariance condition looks like
$$a_0g'_1(x)+a^2_0(g^2_1+g'_2(x))=-\Bigl({q a_0\over
 1+pa_0}\Bigr) g'_1(x)+\Bigl({q a_0\over 1+pa_0}\Bigr)^2
(g^2_1-g'_2(x))+R_1a_0+R_2a^2_0.\eqno{(7.11)}$$
Expanding the denominators in powers of $a_0$ and equating terms
of order $a_0$ and $a^2_0$ yields equations for the functions $g_1(x)$ and
$g_2(x)$ which give
$$g_1(x)={R_1 x\over 1+q}, \qquad g_2(x) = \bigg [ R_2+({pqR_1\over
(1+q)})\bigg ]x-{(1-q)\over (1+q)^2(1+q^2)}{x^3\over 3}.\eqno{(7.12)}$$
The energy spectrum which follows from eqs.(3.11), (3.15) and (7.9)
is $E_0^-=0$ and
$$E^{(-)}_n=R_1\sum^n_{j=1}{q^{j-1}a_0\over [1+pa_0({1-q^{j-1})\over 1-q})]}
+R_2\sum^n_{j=1}{(q^{j-1}a_0)^2\over [1+pa_0({1-q^{j-1})\over (1-q)})]^2},
	\qquad n=1,2,\ldots \eqno{(7.13)}$$
As usual, $\psi^{(-)}_n$ and $T_-$ can be found using eqs.(3.5), (3.27), (3.30)
and
(3.31). Generalization to the case when arbitrary numbers of the $R_j$ are
nonzero is straightforward. Similarly, one can also consider shape
invariance in multi-steps along with the ansatz (7.9) and obtain
deformations of the multi-soliton Rosen-Morse solutions.
\vskip 0.5 true cm
\noindent {\bf VIII. Summary and Open Problems}

Until now, the only known shape invariant potentials were such that
the parameters $a_1$ and $a_0$ which
appear in shape invariance condition (1.1) were related by a
translation. In this paper, we have discovered a wider class of
new shape invariant potentials
for which $a_1$ and $a_0$ are related by scaling, as well as in
a variety of other ways. All these new potentials are
reflectionless and have an infinite number of bound states.
They can be considered to be q-deformations of the multi-soliton solutions
corresponding to the Rosen-Morse potential. We were able to  obtain the
energy eigenvalues, eigenfunctions and transmission coefficients
for these potentials algebraically.
It was also possible to obtain
analytical answers for the moments of these potentials, which should
be useful since these potentials could not be explicitly expressed in a
closed analytic form.
The recently discovered self-similar
potentials of Shabat and Spiridonov were shown to be a very special case of our
shape invariant potentials. We were also able to obtain q-deformations
of the one
dimensional harmonic oscillator potential. This work has raised
several questions which need to be looked into. Some of these are:
\item {(i)} Just as we have obtained q-deformations of the
reflectionless Rosen-Morse and harmonic oscillator potentials, can
one also obtain deformations of the other simple shape invariant potentials?
In particular, can one obtain deformations of potentials
which are not reflectionless, say non-solitonic Rosen-Morse
potentials of the form $V(x) = -A(A+1) \hbox{sech}^2 x$ for non-integer
values of $A$.
\item {(ii)} What are the various potentials satisfying the shape
invariance condition (3.9)?. In this paper, we have significantly
expanded that list but it is clear that the possibilities are far from
exhausted. In fact it appears that there are an unusually large number
of shape invariant potentials, for all of which the whole
spectrum can be obtained algebraically. How does one classify all
these potentials? Would such a classification exhaust all the known exactly
solvable ones dicussed by Natanzon [19] ?
\item {(iii)} The shape invariant potentials have been treated
algebraically in this paper. An obvious interesting question is if
one can also solve the Schr\"odinger equation for these potentials directly?
This
should be possible, at least in principle. In that case the next
question is if the Schr\"odinger equation gets essentially reduced to
a hypergeometric or confluent hypergeometric equation or not. If not,
then one would have generalized the concept of solvable potentials as
introduced by Natanzon [19].
\item {(iv)} Now that a host of new shape invariant potentials have
been discovered,
it is worth asking if all the known exactly solvable potentials of
Natanzon can be cast in a shape invariant form. In fact one can
ask an even more general question: can any exactly solvable problem in
quantum mechanics (i.e. for which the Schr\"odinger equation need not
necessarily reduce to a hypergeometric or confluent hypergeometric
equation) be cast in a shape invariant form? In other words, is shape
invariance not only sufficient but even necessary for exact
solvability, as first conjectured by Gendenshte\u \i n [6]?

We hope to answer some of these questions in the future.
\vskip 0.5 true cm
\noindent {\bf Acknowledgements}

One of us [U.S.] would like to thank the Council of Scientific and
Industrial Research, India and the United Nations Development Programme for
support under their TOKTEN scheme and the kind hospitality of the
Institute of Physics, Bhubaneswar where a part of this work was done.
This work was supported in part by the U.S. Department of Energy.
\vfill
\eject
\noindent {\bf References}
\item {[1]} A. Shabat, Inverse Prob. {\bf 8}, 303 (1992).
\item {[2]} V.P. Spiridonov, Phys.Rev.Lett.{\bf 69}, 298 (1992).
\item {[3]} V.P. Spiridonov, Univ. de Montreal preprint UdeM-LPN-TH94-92.
\item {[4]} V.A. Rubakov and V.P. Spiridonov, Mod.Phys.Lett.{\bf A3}, 1337
(1988);
\item {} A. Khare, J.Phys.{\bf A 25}, L749 (1992); J.Math.Phys. (1993), in
press.
\item {[5]} E. Witten, Nucl.Phys.{\bf B185}, 513 (1981);
\item { } F. Cooper and B. Freedman, Ann.Phys.(N.Y){\bf 146}, 262 (1983).
\item {[6]} L. Gendenshte\u \i n, JETP Lett.{\bf 38}, 756 (1983).
\item {[7]} R. Dutt, A. Khare and U.P. Sukhatme, Phys.Lett.{\bf
181B}, 295 (1986);
\item { }  J. Dabrowska, A. Khare and U.P. Sukhatme, J.Phys.{\bf A21}, L195
(1988).
\item {[8]} A. Khare and U.P. Sukhatme, J.Phys.{\bf A21}, L501 (1988);
\item { }  W.Y. Keung et al, J.Phys.{\bf A22}, L987 (1989).
\item {[9]} R. Dutt, A. Khare, U.P. Sukhatme, Am.J.Phys.{\bf 56},
163 (1988);
\item {} L. Infeld and T. Hull, Rev.Mod.Phys.{\bf 23}, 21 (1951).
\item {[10]} F. Cooper, J.N. Ginocchio and A. Khare, Phys.Rev.{\bf
D36}, 2458 (1987).
\item {[11]} D.T. Barclay and C.J. Maxwell, Phys.Lett.{\bf A157},
357 (1991).
\item {[12]} A short summary of this work has been given in A. Khare and
U.P. Sukhatme, Inst. of Physics, Bhubaneswar preprint IP/BBSR/92-93.
\item {[13]} A. Degasperis and A. Shabat, Univ. di Roma preprint n.919, 1992.
\item {[14]} See for example G.L. Lamb, Elements of Soliton Theory,
Wiley (N.Y) 1980 or A. Das, Integrable Models, World Scientific,
Singapore, 1989.
\item {[15]} W.Y. Keung, E. Kovacs and U.P. Sukhatme,
Phys.Rev.Lett.{\bf 60}, 41 (1988).
\item {[16]} J.F. Schonfeld, W. Kwong, J.L. Rosner, C. Quigg and H.B.
Thacker, Ann.Phys.(N.Y){\bf 128}, 1 (1980) and references therein.
\item {[17]} L.C. Biedenharn, J.Phys.{\bf A22}, L873 (1989);
\item {  } A.M. Macfarlane, J.Phys.{\bf A22}, 4581 (1989).
\item {[18]} M.M. Nieto, Phys.Lett.{\bf 145B}, 208 (1984).
\item {[19]} G.A. Natanzon, Vestnik Leningrad Univ.{\bf 10}, 22 (1971);
Teoret.Mat.Fiz.{\bf 38}, 146 (1979);
\item { } see also J.N. Ginocchio, Ann.Phys.(N.Y){\bf 152}, 203 (1984); {\bf
159}, 467 (1985).
\vfill
\eject
\noindent {\bf Figure Captions}
\item {Figure 1} Self-similar superpotentials $W(x)$ for various
values of the deformation parameter $p$. The Taylor series are summed
for $x<0.4$ and the functions extrapolated from there by numerically
solving  the self-similarity condition, eq.(2.9).
\item {Figure 2} Self-similar potentials $V_-(x)$ (symmetric about $x=0$)
corresponding to the superpotentials graphed in Figure 1.
\item {Figure 3} A double well potential $V_-(x)$ (solid line)
and its shape invariant, single well supersymmetric partner $V_+(x)$
(dotted line). The exact spectra
are also displayed. Parameter values are $r_1=1$, $r_2=-1$, $q=0.3$
and $a=0.75$.
\item {Figure 4} Selected members of the one parameter family of
isospectral potentials which includes the self-similar potential with $p=0.5$.
Note that a different choice of $C_1$ (i.e. $r_1$) has been made
compared to Figures 1 and 2.

\vfill
\eject
\end